\documentclass[conference]{IEEEtran}
\IEEEoverridecommandlockouts

\usepackage{cite}

\usepackage{amssymb,amsfonts,amsthm}
\usepackage[fleqn]{amsmath}

\usepackage{algorithmic}
\usepackage{graphicx}
\usepackage{textcomp}
\usepackage{xcolor}

\usepackage{algorithm}

\usepackage{graphicx}
\usepackage{caption}
\usepackage{subcaption}
\usepackage{url}

\usepackage{balance}

\usepackage{mdframed}

\def\BibTeX{{\rm B\kern-.05em{\sc i\kern-.025em b}\kern-.08em
    T\kern-.1667em\lower.7ex\hbox{E}\kern-.125emX}}

\usepackage{todonotes}

\DeclareCaptionLabelFormat{andtable}{#1~#2  \&  \tablename~\thetable}

\begin{document}

\title{Adaptive Neural Architectures \\for Recommender Systems}

\author{\IEEEauthorblockN{Dimitrios Rafailidis}
\IEEEauthorblockA{\textit{Maastricht University} \\
Maastricht, The Netherlands \\
dimitrios.rafailidis@maastrichtuniversity.nl}
\and
\IEEEauthorblockN{Stefanos Antaris}
\IEEEauthorblockA{\textit{KTH Royal Institute of Technology} \\
Stockholm, Sweden \\
antaris@kth.se}
}

\maketitle

\begin{abstract}
Deep learning has proved an effective means to capture the non-linear associations of user preferences. However, the main drawback of existing deep learning architectures is that they follow a fixed recommendation strategy, ignoring users' real time-feedback. Recent advances of deep reinforcement strategies showed that recommendation policies can be continuously updated while users interact with the system. In doing so, we can learn the optimal policy that fits to users' preferences over the recommendation sessions. The main drawback of deep reinforcement strategies is that are based on predefined and fixed neural architectures. To shed light on how to handle this issue, in this study we first present deep reinforcement learning strategies for recommendation and discuss the main limitations due to the fixed neural architectures. Then, we detail how recent advances on progressive neural architectures are used for consecutive tasks in other research domains. Finally, we present the key challenges to fill the gap between deep reinforcement learning and adaptive neural architectures. We provide guidelines for searching for the best neural architecture based on each user feedback via reinforcement learning, while considering the prediction performance on real-time recommendations and the model complexity.

\end{abstract}

\begin{IEEEkeywords}
Adaptive neural models, recommender systems, deep reinforcement learning
\end{IEEEkeywords}

\section{Introduction}

Following the collaborative filtering strategy, latent models such as matrix factorization~\cite{SalakhutdinovM07} and factorization machines~\cite{Rendle10} have been widely used to generate personalized recommendations. To capture the non-linearity in user preferences, several deep learning strategies have been introduced ~\cite{Eb18,p22,He17,p23}. However, these models follow a fixed recommendation policy which does not correspond to the dynamic real-world scenario, as in practice users evolve their preferences while interacting with recommender systems~\cite{BonnerV18,HidasiKBT15}. Recommender systems, instead of having a static recommendation strategy, should continously update their policies according to users' real-time feedback. For instance, in real-time recommendations a user may want to seek for alternatives with diverse topics of interest, requiring to adapt the recommendation policy in real-time~\cite{ZhengZZXY0L18}.

To capture users' drift on their preferences several time-aware latent models have been introduced~\cite{Koren10,RafailidisN15a}. However, such models aim to learn users' long-term preferences. Instead, sequential recommender systems try to maximize the immediate reward for future recommendations, that is to predict the next recommendations for the short-term sessions~\cite{HidasiK18,HidasiKBT15,LiuZMZ18,WangCZLL18}. However, sequential recommendations fail to accurately predict the long-term rewards in the future~\cite{Zhao18}. To handle this issue, some attempts have been made trying to search for both short-term and long-term rewards via Reinforcement Learning (RL), by finding the best recommendation policy based on an action-value function~\cite{HuDZ0X18,MahmoodR09}. The main challenge in baseline RL methods like Q-Learning~\cite{TaghipourK08} and Partially-Observed Markov Decision Process (POMDP)~\cite{KearnsMN02} is the extremely large number of possible items in recommender systems, resulting in enormous state and action spaces. To face this problem, more recently a few studies exploit deep RL to approximate the action-value function of baseline RL methods, thus supporting huge amount of items in recommender systems~\cite{Zhao18,ZhaoZDXTY18,ZhengZZXY0L18}. Based on the neural architecture of the Deep Q-Network (DQN) model~\cite{LillicrapHPHETS15,MnihKSGAWR13}, deep RL methods can continuously update their recommendation policies based on users' real-time feedback. Although deep RL methods in recommender systems try to learn the optimal strategy that fits users' preferences, they rely on predefined and fixed neural architectures with high complexity. In doing so, they cannot dynamically adjust their neural architectures over the user sessions based on the achieved performance by the recommendation strategy~\cite{Yang19}.  

Accounting for the importance of adaptive neural architectures in recommender systems, the main contribution of this study is summarized as follows:
\begin{itemize}
\item \emph{We investigate the main limitation of recent state-of-the-art reinforcement learning strategies in recommender systems, which are built on fixed neural architectures over the recommendation sessions.} 
\item \emph{In addition, we present how adaptive neural architectures are developed for consecutive tasks in other research domains.}
\item \emph{To bridge this gap we discuss the key challenges to produce recommendations by dynamically adapting neural architectures via deep reinforcement learning.}
\end{itemize}

\section{Reinforcement Learning in Recommender Systems} \label{sec:rel}
Several attempts have been made to generate recommendations in real-time via RL, assuming that users keep interacting with the recommender system. For example, Taghipour et al.~\cite{TaghipourK08} formulate a Q-Learning problem and learn to generate recommendations based on user feedback on the Web. This approach inherits the intrinsic characteristic of RL, performing a constant learning process. Mahmood et al.~\cite{MahmoodR09} adopt the RL technique to observe the responses of users in conversational recommender systems. The goal is to maximize a cumulative reward function which corresponds to the benefit that users receive from a recommendation session. 

Recently, Zheng et al.~\cite{ZhengZZXY0L18} propose a RL framework to generate online news recommendation. Bonner and Flavian~\cite{BonnerV18} introduce a RL algorithm to learn from logged data based on a biased recommendation policy. This algorithm tries to compute the best recommendation policy for maximizing the reward by taking into account the control recommendation policy for each user. Zhao et al.~\cite{Zhao18} propose a deep RL strategy for generating page-wise recommendations based on the actor-critic framework~\cite{SuttonB98}. This approach can optimize a page of items with proper display based on real-time feedback from users. Bai et al.~\cite{Bai19} present a self-attentive recommendation model for capturing the evolving demands of users over time focusing on long-term demands e.g., repeated purchasing with a persistent interest and short-term demands e.g., buying the complementary purchasing in a short time period. However, this study does not aim at generating long-term novel recommendations but repeated ones. Gui et al.~\cite{GUI19} introduce a cooperative multi-agent reinforcement learning model for mention recommendation on Twitter. Shang et al.~\cite{Shang19} adopt a multi-agent generative adversarial reinforcement learning framework for recommendations on a ride-hailing platform. Zou et al.~\cite{Zou19} introduce a deep RL framework for short and long-term click prediction. However, this framework does not produce personalised recommendations by focusing on the click prediction problem. Zhao et al~\cite{ZhaoZDXTY18} present DEERS, a pairwise deep RL framework. In this work, the best strategies are computed via recommending trial-and-error items and receiving reinforcements of these items from both positive and negative user feedback. Lei et al.~\cite{LeiPY020} demonstate how to use a RL agent for a Graph Convolutional Q-network, aiming to compute recommendation policies based on the graph-structured representations. Xin et al.~\cite{XinKAJ20} propose a self-supervised RL strategy for sequential recommendations. Hong et al.~\cite{HongLD20} introduce a music recommendation framework for adapting to user's current preference based on reinforcement learning in real time during a listening session. Zhao et al.~\cite{ZhaoZZZBY20} present a deep hierarchical reinforcement learning  framework to capture the long-term sparse conversion interest at the high level and automatically set abstract goals. At the the low-level of the hierarchy, the learning strategy tries  to meet the abstract goals and model short-term click interest when interacting with a real-time environment. 

In addition, Zhou et al.~\cite{ZhouDC0RTH020} investigate how to exploit knowledge graphs (KG) for RL in recommender systems. Instead of learning RL policies from scratch, this approach first exploits the prior knowledge of the item correlation from KG to guide the candidate selection, enriches the representation of items and user states, and then propagates user preferences among the correlated items to handle the sparsity in user preferences. In a similar spirit, Wang et al.~\cite{WangFXZNH20} generate sequential recommendations based on RL and KG. Zhao et al.~\cite{ZhaoWZZLX020} supervise the path finding process in KG via an RL framework to produce explainable recommendations. 

\emph{ The main limitation of the aforementioned state-of-the-art RL strategies in recommender systems, is that the neural architectures are fixed during training. Consequently, they do not focus on the recommendation performance based on the user real-time feedback and do not adapt the neural architectures over the recommendation sessions in real-time, accordingly. This might significantly limit the performance of deep RL strategies when producing recommendations, as for example occurs in the case of consecutive tasks in other research domains~\cite{ZacariasA18,ZenkePG17,YoonYLH18,RusuRDSKKPH16,ShenZWJ18}.}

\section{Adaptive Neural Networks}
Instead of having a fixed neural architecture during training, recent studies introduce models with the ability to learn consecutive tasks, such as natural language processing~\cite{Has17}, speech synthesis~\cite{Wu15}, image segmentation~\cite{Heg17}, object detection~\cite{Red17} and domain independent feature decomposition~\cite{XuYY020} in the cross-domain retrieval task~\cite{RafailidisC16}. The goal of such models is to solve new tasks, without forgetting the acquired knowledge from previous tasks, also known as lifelong learning in neural architectures~\cite{ParisiKPKW19}. The main idea is for every new task to dynamically augment the neural architecture while preserving the parameters of the previous architecture unchanged. In particular, provided that in the real-world setting we have to train the neural network in consecutive tasks, in the learning  strategy of the neural network we have to introduce regularization terms to preserve the model parameters similar to the learned parameters of the previous sessions, by considering in each new session their importance for each task~\cite{XuZ18,ZacariasA18,ZenkePG17}. 

A few studies explore neural networks capable of dynamically adjusting the ``capacity'' of the neural architecture, that is the numbers of the hidden layers and units when training the neural network. For example, Zhou et al.~\cite{ZhouSL12} perform  incremental learning for a denoising autoencoder by adding new neurons for a certain group of ``difficult'' training examples. This is achieved by setting high loss for each group of ``difficult'' examples, and then combine them with other neurons to reduce unecessary complexity of the neural network and avoid redundancy. Provided a fixed size of hidden units and layers, Progressive Neural Networks (PNN)~\cite{RusuRDSKKPH16,ShenZWJ18} expand the neural architecture. In practice though, this expansion strategy in PNN results in a large network structure for many sequential tasks, as for each new task the PNN stucture is significantly augmented. This means that in the PNN model the network architecture might become extremely large. This problem becomes even more challenging in the case of recommender systems where users continously interact to provide the system with feedback in different forms such as ratings, views, clicks and so on. The deep learning strategy of Dynamically Expandable Network (DEN)~\cite{YoonYLH18} tries to face the problem of high complexity by setting group sparsity regularization terms to the newly added parameters in each new session. Nonetheless, in practice this strategy introduces many hyperparameters in DEN. As a consequence, this results in several regularization and thresholding hypeparameters, requiring a lot of effort to tune the model and improve its performance. 

Xu et al.~\cite{XuZ18} propose a reinforced continual learning framework for the image classification task. This framework adapts the neural architecture by adjusting the numbers of hidden units and layers, accordingly. The reinforced continual learning framework consists of the following three networks: the controller, the user-session network, and the value network. The controller is  a Long Short-Term Memory network (LSTM) for generating policies and computing how many hidden units and layers will be added in each session. The user-session network is an extension of the DQN model by adjusting the neural architecture according to the performance accuracy of a reinforcement agent. The value network is designed as a fully-connected network, which approximates the value of the state per session. 

\emph{Different from baseline strategies with dynamically adjusted neural architectures which mainly focus on classification tasks, recommendation strategies have to adapt the underlying neural architecture by capturing users' preference dynamics over the sessions to boost the recommendation accuracy in a personalized manner.} 

\section{Key Challenges}
Recently, deep RL strategies have been widely used in recommender systems to capture user feedback in real time and generate recommendation strategies. However, there is a still a technological gap between baseline deep RL strategies with fixed neural architectures and adaptive ones developed in other research domains. In addition, we have to account for the fact that neural architectures are not always the most suitable learning strategy in recommender systems, as pointed out in~\cite{DacremaCJ19}. 

The key challenges to develop a deep RL strategy with an adaptive neural architecture are summarized as follows:

\begin{itemize}
\item \emph{When new user feedback/data arrive in a recommendation session, how can we formulate an optimization problem to decide the optimal number of hidden units and/or layers in a  existing neural architecture? In particular, the main challenge is to find the optimal way to configure the neural network in each recommendation session, considering both users long-term and short-term preferences.} 
\item \emph{In deep RL strategies, the challenge is to find how to model a reward signal by taking into account both the recommendation accuracy and the neural architecture's complexity over the user sessions.}
\item \emph{Provided that deep RL strategies require the computation of a high number of model parameters, how can we develop an adaptive strategy to adjust the neural architecture based on the recommendation performance over the user sessions, while at the same time preserving the model complexity low? In the real-world setting, continuous augmentation of an existing neural architecture might result to prohibited training cost. This means that researchers have to be careful in the expansion strategy of the neural architecture when new user data arrive in a recommendation session, so as to preserve a medium-size network.}
\item \emph{Which are reliable indicators to evaluate the trade-off between quality of recommender systems based on deep RL strategies and the model complexity of an adaptive neural architecture?} 
\end{itemize}

Designing an adaptive neural architecture in an RL framework can be beneficiary in recommender systems. However, this is a challenging task where researcher have to take into account several factors, such as preserving the model complexity low and adjusting the neural architecture according to users' preference dynamics in the recommendation sessions.

\balance
\bibliographystyle{unsrt}
\bibliography{IEEEfull,references}

\end{document}